\documentclass[10pt,conference]{IEEEtran}
\IEEEoverridecommandlockouts

\usepackage{orcidlink}
\usepackage{cite}
\usepackage{amsmath,amssymb,amsfonts}
\usepackage{algorithmic}
\usepackage{graphicx}
\usepackage{textcomp}
\usepackage{subcaption}
\usepackage{booktabs}
\usepackage{ragged2e}
\usepackage[table]{xcolor}
\usepackage{tabularx}
\usepackage{tablefootnote}
\usepackage{footnote}
\usepackage{multirow}
\usepackage{multicol}
\newcolumntype{P}[1]{>{\RaggedRight\hspace{0pt}}p{#1}}
\usepackage{hyperref}
\usepackage{booktabs}
\usepackage{tikz}
\usetikzlibrary{positioning}
\usepackage{url}
\usepackage[most]{tcolorbox}
\tcolorboxenvironment{rqbox}{
  colback=gray!10,
  colframe=gray!40,
  boxrule=0.3pt,
  arc=2pt,
  left=6pt,
  right=6pt,
  top=4pt,
  bottom=4pt
}

\hypersetup{
    colorlinks=true,
    linkcolor=black,
    filecolor=black,      
    urlcolor=black,
    citecolor=black,
}

\definecolor{tableGray}{RGB}{243, 244, 245}
\definecolor{blueshade}{RGB}{231, 238, 243}
\definecolor{lightgrey}{RGB}{200,200,200}

\newtcolorbox{boxA}{
    colback = blueshade, 
    boxrule = 0pt  
}

\newtcolorbox{boxB}{
    colback=tableGray, colframe=lightgrey, boxrule=0.5pt, arc=2mm, left=5pt, right=5pt, top=5pt, bottom=5pt
}

\def\BibTeX{{\rm B\kern-.05em{\sc i\kern-.025em b}\kern-.08em
    T\kern-.1667em\lower.7ex\hbox{E}\kern-.125emX}}

\begin{document}

\title{AI Transparency: Governance Compliance or Stakeholder Requirements?}

\author{\IEEEauthorblockN{Muneera Bano \orcidlink{0000-0002-1447-9521}}
\IEEEauthorblockA{\textit{CSIRO} \\
Melbourne, Australia \\
muneera.bano@csiro.au }
\and
\and
\IEEEauthorblockN{Didar Zowghi \orcidlink{0000-0002-6051-0155}}
\IEEEauthorblockA{\textit{CSIRO} \\
Sydney, Australia \\
didar.zowghi@csiro.au}
}

\maketitle

\begin{abstract}
Transparency is increasingly mandated for public-sector AI systems, with organisations required to publish statements describing their AI use and oversight arrangements. However, the existence of such artefacts is often treated as equivalent to transparency itself, despite limited evidence that they proportionately serve relevant stakeholder groups. From a requirements engineering perspective, this raises a validation concern: compliance with mandated disclosure criteria does not necessarily ensure transparency adequacy for stakeholders with different levels of risk exposure, decision control, and involvement. This paper presents an empirical analysis of 92 publicly available AI transparency statements published by Australian Government agencies under the national AI governance mandate. We introduce the stakeholder Risk--Control--Involvement--Need (RCIN) framework to differentiate stakeholder classes according to their structural position and transparency needs. Using a structured rubric derived from the mandated criteria, we evaluate how both the mandate and published statements are calibrated to each stakeholder class. The findings show that while structural compliance is widespread, transparency calibration is uneven. Criteria serving high-control stakeholders are consistently realised, whereas criteria most critical for high-risk, low-control stakeholders are fewer and less substantively addressed. We conceptualise this as the Transparency Illusion: a condition in which transparency appears satisfied through compliant artefacts yet remains unevenly calibrated to stakeholders bearing the greatest exposure to AI-supported decisions. The study frames transparency as a stakeholder-calibrated validation problem, demonstrating that artefact-level compliance does not constitute requirements validation in this context.
\end{abstract}

\begin{IEEEkeywords}
Requirements Engineering; Artificial Intelligence; Transparency; AI Governance
\end{IEEEkeywords}

\section{Introduction}

Artificial intelligence (AI) systems are increasingly deployed by public and private sector organisations to support decision-making, service delivery, and administrative 
processes \cite{hjaltalin2024strategic, van2022artificial}. As these systems 
influence people's rights and access to services, transparency has become a 
central principle of responsible AI governance \cite{lu2023responsible}, framed 
not merely as a desirable system quality but as a mandated governance obligation 
\cite{Felzmann2020Towards, Balasubramaniam2022Transparency, Wang2025Signaling}. 
Major governance regimes, including the OECD AI Principles \cite{OECD2019AI}, 
the NIST AI Risk Management Framework \cite{NIST2023AI}, the World Economic 
Forum's AI governance guidance \cite{WEF2023AI}, and the EU AI Act 
\cite{EUAIAct2024}, consistently position transparency as foundational to 
accountability, oversight, and trust. 

In practice, transparency is operationalised through documentation and disclosure 
artefacts rather than as an intrinsic system property. Organisations publish 
transparency statements as evidence of policy compliance \cite{Wang2025Signaling, ICO2023Transparency}, however, the existence of such artefacts is often treated as 
synonymous with transparency itself, despite limited evidence that they benefit all the stakeholders they are intended to serve effectively \cite{Balasubramaniam2022Transparency, Wang2025Signaling, McCormack2025comprehensive}. 

Although transparency and explainability are often discussed together \cite{Miller2019Explanation, lu2023responsible}, they are not synonymous. Explainability is about explanation of how AI systems generate outputs \cite{Arrieta2020XAI, girard2024inclusive}, whereas transparency also includes organisational disclosure, accountability structures, and contestability \cite{Ananny2018Seeing}. This study examines transparency as a broader governance and requirements engineering concern operationalised through public disclosure artefacts rather than only through technical explanation mechanisms \cite{Larsson2020Transparency}.

From a requirements engineering (RE) perspective, transparency is typically conceptualised as a non-functional requirement \cite{chung2012non, Larsson2020Transparency, Ahmad2021Whats}. However, RE distinguishes between requirements \emph{satisfaction}, conformance to specified criteria; and \emph{validation}, confirmation that requirements fulfil stakeholder needs in context. Prior work emphasises that transparency is relational rather than intrinsic, depending on what and how information is communicated to and interpreted by stakeholders \cite{Chen2021INTRPRT}. 

Stakeholder-oriented approaches argue for participatory engagement and differentiated transparency mechanisms tailored to stakeholder needs \cite{Bell2023Think, Chaudhry2022Transparency, 
Paterakis2025Comprehensive}. However, governance practice frequently evaluates 
transparency through structural compliance checks rather than stakeholder-oriented 
validation. Distinct stakeholder classes such as: rule-setting authorities, system implementers, organisational deployers, and affected individuals; differ substantially in their AI risk exposure, decision control, and accountability roles. Research demonstrates that power asymmetries influence which stakeholders' requirements are prioritised and how effectively they are realised \cite{bano2015systematic, bano2017user, bano2018power}. However, governance mandates prescribe uniform disclosure formats, 
implicitly assuming a single artefact can satisfy diverse and competing 
transparency needs \cite{Wang2025Signaling, Zowghi2025On}.

We address this gap through an empirical study of 92 publicly available AI transparency statements published by Australian public-sector organisations under the national policy for AI in government. Treating transparency statements as requirements-related artefacts, we assess whether disclosure patterns align with stakeholder needs across risk, control, and involvement dimensions. To the best of our knowledge, no prior empirical study has systematically evaluated governance-mandated transparency artefacts against diverse stakeholder requirements. The study addresses two main research questions: \\

\begin{itemize}
    \item \textbf{RQ1:} \textit{How are the mandated AI transparency criteria structured for relevant stakeholder classes?}
    \item \textbf{RQ2:} \textit{To what extent do published AI transparency statements meet the transparency requirements of  relevant stakeholder classes?}
\end{itemize}

RQ1 examines the structure of the mandate itself, while RQ2 evaluates how published transparency statements realise those requirements in practice. The paper makes three main contributions:

\begin{itemize}
    \item \textbf{A stakeholder-calibrated transparency framework.} We introduce 
    the stakeholder Risk--Control--Involvement--Need (RCIN) framework, a 
    theoretically grounded analytical lens for evaluating AI transparency 
    requirements according to stakeholder-specific risk exposure, decision control, 
    degree of involvement, and informational need. This framework conceptualises transparency adequacy as contingent on stakeholder calibration rather than uniform 
    disclosure compliance.

    \item \textbf{A structural evaluation of mandated transparency statements.} We present the first large-scale empirical analysis of 92 publicly available AI transparency statements, examining how mandated criteria distribute transparency obligations across stakeholder classes and whether published statements meet the informational requirements of all relevant stakeholder classes.
    
  \item \textbf{The Transparency Illusion as a requirements validation challenge.} We define and analyse the concept of \emph{Transparency Illusion}: a structural condition in which transparency is nominally satisfied through compliance-driven artefacts yet remains misaligned with the needs of high-risk, low-control stakeholders. This re-frames AI transparency within RE not as a documentation or disclosure challenge, but as a validation problem concerned with whose requirements are recognised, prioritised, and meaningfully served.

\end{itemize}

The remainder of the paper is structured as follows. Section~\ref{sec:relatedwork} 
reviews related work. Section~\ref{sec:framework} introduces the RCIN framework. 
Section~\ref{sec:method} describes the research design. Section~\ref{sec:results} 
presents results, followed by discussion, threats to validity, and conclusions.

\section{Background and Related Work}
\label{sec:relatedwork}

In this section, we examine how transparency is conceptualised in the literature, how it is addressed in AI governance frameworks, and how it is operationalised as a requirement. We then synthesise the findings from literature to present the gap motivating this study.

\subsection{AI Transparency}

The literature reveals substantial heterogeneity in how AI transparency is defined 
and framed. It is frequently conceptualised as a multidimensional construct 
encompassing traceability, explainability, communication, accessibility, and 
disclosure \cite{Felzmann2020Towards, Balasubramaniam2022Transparency}. It is also treated as a non-functional requirement shaping system quality 
\cite{Mancine2024Estado} or as a foundational component of trustworthy AI alongside 
accountability and fairness \cite{McCormack2025comprehensive, Nastoska2025Evaluating}. 
A significant theoretical contribution conceptualises transparency as relational 
rather than intrinsic: Chen et al. describe it as an 'affordance' between 
algorithm and user, emphasising that transparency depends on stakeholder 
interpretation rather than solely on technical properties \cite{Chen2021INTRPRT}. 
This relational perspective is echoed in interdisciplinary analyses distinguishing 
transparency as a normative value, a technical mechanism, and a socio-technical 
relation \cite{Felzmann2020Towards, Langer2021What}.

In Responsible AI literature, transparency is closely associated with explainability, accountability, and trust, but these concepts are not synonymous. Explainability concerns understanding how AI systems generate outputs or decisions \cite{Arrieta2020XAI, girard2024inclusive}. Transparency further includes organisational disclosure, accountability structures, procedural visibility, and mechanisms enabling stakeholder oversight and contestability \cite{Ananny2018Seeing}. Explainability may support transparency and trust, but does not fully constitute them. This distinction is particularly important in governance contexts, where transparency is often operationalised through documentation, reporting obligations, and public disclosure artefacts rather than through technical explanation mechanisms alone. Recent systematic reviews highlight definitional divergence and conceptual fragmentation across the field 
\cite{Nauta2023From, Cao2025Systematic, kemmerzell2025towards}. Despite widespread endorsement of 
transparency as a core AI ethics principle, consensus on its dimensions, evaluation 
criteria, and stakeholder alignment remains limited, complicating its 
operationalisation in practice.

\subsection{AI Transparency in Governance Frameworks}

In governance contexts, transparency is institutionalised as both a normative 
principle and a compliance obligation, though its form and scope vary across 
jurisdictions. The EU AI Act establishes differentiated obligations according to 
system risk classification, subjecting high-risk systems to structured documentation 
requirements covering system purpose, design, data governance, risk management, and 
human oversight \cite{Nannini2024Operationalizing}. In the 
United States, the NIST AI Risk Management Framework formalises transparency as a 
core trustworthiness characteristic, emphasising documentation, traceability, and 
communication across the AI lifecycle \cite{NIST2023AI}. At the global level, the 
OECD AI Principles articulate transparency as a shared normative commitment 
proportionate to system context and risk \cite{OECD2019AI}.

In Australia, the AI Ethics Principles identify transparency and explainability as 
core requirements, stating that individuals should be informed when AI systems 
significantly affect them \cite{AustraliaEthics2019}. The Australian Government additionally 
mandates public AI transparency statements for agencies deploying AI systems 
\cite{DTA2025Transparency}, formalising transparency as a reporting obligation 
linked to accountability and institutional legitimacy \cite{Wang2025Signaling}. Wang \cite{Wang2025Signaling} cautions, however, that transparency may operate symbolically, serving compliance 
and reputational purposes rather than substantively empowering affected stakeholders.

Across jurisdictions, transparency is consistently framed as essential for accountability and trust, however, its operationalisation remains predominantly artefact-driven. Governance frameworks prescribe disclosure structures but rarely specify how transparency should be validated against stakeholder requirements, risk exposure, decision control, or degree of involvement.

\subsection{AI Transparency as a Requirement}

Several frameworks, rubrics, and evaluation mechanisms for operationalising transparency requirements have been proposed. Transparency by design articulates structured principles spanning system design, information provision, and accountability \cite{Felzmann2020Towards}. The HXAI framework integrates 
transparency throughout the machine learning lifecycle \cite{Paterakis2025Comprehensive}, 
while the XAIR metareview aligns explainability mechanisms with software development 
lifecycle stages \cite{Clement2023XAIR}. Checklist-based instruments include 
Transparency-Check \cite{Schelenz2020Applying, Schelenz2024Transparency}, audit 
cards, model cards, and datasheets \cite{Staufer2025Audit, Hutchinson2021Towards}, 
and maturity models that stage organisational transparency capability 
\cite{Munoz2025Maturity, Kioskli2025trustSense}.

From an RE perspective, stakeholder-first elicitation and participatory design 
approaches are proposed to capture transparency needs \cite{Bell2023Think}, and 
structured specification mechanisms with lifecycle integration strategies have been 
advocated \cite{Li2023Dealing, Mancine2024Estado}. However, traceability and 
post-deployment validation of transparency requirements remain underdeveloped 
\cite{Villamizar2021Requirements, Clement2023XAIR}. Research on stakeholder 
involvement demonstrates that power asymmetries influence whose requirements are 
prioritised and how satisfaction is experienced \cite{bano2015systematic, 
bano2017user, bano2018power}, raising questions about whether governance-driven 
transparency artefacts adequately reflect the needs of those with the highest risk 
exposure but the least influence over system design.

\subsection{Research Gap}

Despite significant advances, empirical validation of governance-mandated 
transparency artefacts in real-world settings remains limited, with most proposed 
frameworks evaluated through small-scale case studies rather than live public-sector 
deployments \cite{Nauta2023From, Cao2025Systematic}. Three interrelated gaps emerge: 
(1) stakeholder inclusion is often constrained, prioritising technical or regulatory 
actors over affected end-users \cite{Langer2021What, Balasubramaniam2023Transparency, bano2018power}; 
(2) compliance-driven documentation does not necessarily translate into meaningful 
stakeholder understanding \cite{Wang2025Signaling}; and (3) systematic evaluation of 
real-world transparency statements as requirements-related artefacts remains 
underexplored \cite{McCormack2025comprehensive, Cao2025Systematic, Balasubramaniam2023Transparency}. This study addresses these gaps by empirically analysing a large corpus of governance-mandated AI transparency statements through a stakeholder Risk--Control--Involvement--Need lens, contributing an evaluation-focused perspective to AI transparency and RE-related research.

\section{Risk--Control--Involvement--Need (RCIN) Framework for AI Transparency}
\label{sec:framework}

The RCIN framework was developed through a three-stage structured synthesis. First, 
we identified foundational constructs from three strands of literature: relational 
accounts of transparency in AI governance \cite{Larsson2020Transparency, 
Chen2021INTRPRT}, stakeholder involvement and power asymmetry in requirements 
engineering \cite{bano2015systematic, bano2018power}, and accountability and risk 
exposure in AI governance \cite{Wang2025Signaling, NIST2023AI, EUAIAct2024}. 
Second, we derived candidate dimensions by mapping shared conceptual concerns across 
these strands, identifying structural factors that recur across documented cases of 
transparency misalignment between governance mandates and stakeholder needs. Third, 
we consolidated these candidates into four analytically distinct dimensions through iterative 
alignment with AI transparency statements as governance artefacts, ensuring each 
dimension captures an aspect of transparency adequacy that the remaining three 
dimensions do not address. Together, these dimensions capture complementary aspects of stakeholder position within AI governance: exposure to potential harm (Risk), authority over transparency specification and decision-making (Control), capacity to participate in transparency processes (Involvement), and the informational or procedural requirements necessary to meaningfully engage with AI-supported decisions (Need). While additional dimensions may be possible in other contexts, these four were selected as a minimal but theoretically grounded structure for analysing stakeholder asymmetries within governance-mandated transparency artefacts.

A common insight drawn from this synthesis is that transparency is not a uniform technical property but a socio-technical requirement whose adequacy is inherently contextual \cite{Felzmann2020Towards, Langer2021What}. Governance mandates frequently 
operationalise transparency through standardised formats, implicitly assuming a 
single artefact can satisfy heterogeneous stakeholder expectations 
\cite{Balasubramaniam2022Transparency, Wang2025Signaling}. The RCIN framework 
addresses this by structuring transparency analysis across four dimensions: exposure 
to risk, degree of decision control, level of stakeholder involvement, and 
stakeholder-specific transparency need. The first three characterise a stakeholder's 
structural position within the AI governance ecosystem; the fourth captures the 
substantive informational and procedural requirements arising from that position. 
Crucially, two stakeholders in structurally similar positions may still require 
fundamentally different transparency information, which is why ``Need" constitutes an 
irreducible fourth dimension rather than a derivable consequence of the first three.

\paragraph{Risk Exposure.}
Transparency is consistently justified as a mechanism for mitigating harm and establishing trust, enabling contestability, and protecting fundamental rights \cite{Wang2025Signaling, 
Felzmann2020Towards}. The NIST AI Risk Management Framework conceptualises AI risks 
as potential adverse impacts including bias, discrimination, privacy violations, and 
loss of recourse \cite{NIST2023AI}, while the OECD AI Principles and EU AI Act adopt 
risk-proportionate approaches requiring stronger transparency obligations for 
higher-impact systems \cite{OECD2019AI, EUAIAct2024}. The Australian government's AI Transparency 
Statement template (developed by the Digital Transformation Agency), operationalises this by requiring disclosure of system purpose, oversight structures, data practices, risk management, and human review mechanisms \cite{DTA2025Transparency}. A stakeholder's proximity to potential harm should directly shape what transparency mechanisms are necessary and sufficient for their protection.

\paragraph{Decision Control.}
Decision control refers to the degree of authority a stakeholder exercises over the 
specification, design, implementation, deployment, or governance of an AI system. Research 
consistently demonstrates that stakeholders with greater structural authority exert 
disproportionate influence over requirement specification and implementation 
priorities \cite{bano2018power, bano2017user}. In AI governance contexts, regulators 
and implementers retain structural control over transparency formats and compliance 
criteria while affected individuals have little or no influence 
\cite{Wang2025Signaling, Langer2021What}. When those who define transparency 
obligations are insulated from the consequences of AI decisions, compliance-driven 
artefacts may reflect institutional priorities rather than the protective needs of 
high-risk stakeholders \cite{Balasubramaniam2022Transparency}, 
risking conflation of institutional compliance with genuine stakeholder validation 
\cite{Zowghi2025On}.

\paragraph{Stakeholder Involvement.}
Stakeholder involvement exists along a continuum from passive information provision 
to active participation in decision-making \cite{bano2015systematic, bano2017user}. 
In AI transparency, involvement is not merely procedural but functional: transparency 
operates as an affordance enabling interpretation, evaluation, and potential action 
\cite{Chen2021INTRPRT, Larsson2020Transparency}. A stakeholder who receives a 
disclosure but lacks the contextual knowledge or procedural pathway to act upon it 
is functionally less involved than the disclosure's existence might suggest 
\cite{Langer2021What, McCormack2025comprehensive}. This dimension captures whether 
transparency mechanisms afford meaningful agency, including the capacity to 
understand, query, and contest AI-supported decisions, rather than mere awareness 
\cite{Bell2023Think, Chaudhry2022Transparency}.

\paragraph{Transparency Need.}
Transparency need refers to the type, depth, and form of information required for a 
stakeholder to effectively interpret, evaluate, or act upon AI-supported decisions 
given their structural position. While risk, control, and involvement describe where 
a stakeholder sits within the governance ecosystem, this need captures what they require 
from transparency mechanisms to fulfil their role or protect their interests. Drawing 
on relational accounts of transparency \cite{Chen2021INTRPRT} and stakeholder-oriented 
elicitation in RE \cite{Bell2023Think, Chaudhry2022Transparency, Mancine2024Estado, zowghi2005requirements}, 
we conceptualise need as shaped by the intersection of risk exposure, control 
authority, and involvement, but not reducible to any one factor alone. For instance, 
an affected citizen and a civil society auditor may both have low decision control 
yet require substantially different transparency: the former needs accessible 
explanation and contestability mechanisms, the latter requires technical traceability 
and compliance evidence \cite{Chaudhry2022Transparency, Nannini2024Operationalizing}.

\subsection{Deriving Stakeholder Classes}

Stakeholder classes are identified through a structural role-based approach grounded 
in RE practice \cite{bano2015systematic}, distinguishing stakeholders by their 
position within the AI transparency rather than enumerating individual 
actors \cite{Villamizar2021Requirements, Ahmad2021Whats}. While additional actor 
types such as civil society organisations and parliamentary oversight bodies may 
warrant finer-grained classification in specific contexts, four structurally distinct 
classes are deemed sufficient for our analysing governance-mandated transparency artefacts:

\begin{itemize}
    \item \textbf{Rule-setting authorities}: policy makers and regulators who define 
    transparency obligations, set compliance criteria, and exercise governance 
    oversight.
    \item \textbf{System implementers}: requirements engineers, data scientists, Machine Learning specialists, and 
    developers who translate transparency requirements into technical and documentary 
    artefacts.
    \item \textbf{Organisational deployers}: managers and administrative 
    decision-makers who determine when, where, and how AI systems are applied within 
    institutional settings.
    \item \textbf{Affected individuals and citizens}: service users and members of 
    the public whose rights, access to services, or life outcomes may be influenced 
    by AI-supported decisions.
\end{itemize}

Together these classes span the full governance chain from obligation-setting to 
consequence-bearing, capturing the asymmetric distribution of transparency-related 
power and risk that motivates this study.

\subsection{Structural Asymmetries and Transparency Adequacy}

The stakeholder classes differ systematically across all four RCIN dimensions. 
Affected individuals bear the highest risk exposure yet possess the lowest decision 
control and most limited involvement in transparency specification 
\cite{Wang2025Signaling, Zowghi2025On}, producing transparency needs centred on 
accessible explanation, impact clarity, and contestability mechanisms 
\cite{Bell2023Think, Chaudhry2022Transparency}. Rule-setting authorities and 
implementers possess high control and involvement but lower direct exposure to 
decision consequences, with needs primarily concerning traceability and compliance 
verification \cite{Nannini2024Operationalizing, Sovrano2022Metrics}. Organisational 
deployers occupy an intermediate position, requiring governance visibility and 
accountability mapping \cite{DTA2025Transparency}.

These differences are mutually reinforcing: high risk exposure without corresponding 
control or involvement produces protective, contestability-oriented needs, while high 
control with low direct risk produces process-oriented, compliance-focused needs. 
Treating transparency as uniform across these asymmetric configurations, as 
standardised artefacts implicitly do, produces systematic misalignment between 
governance compliance and stakeholder satisfaction 
\cite{Balasubramaniam2022Transparency, McCormack2025comprehensive}.

\begin{figure}[t]
\centering
\includegraphics[width=\columnwidth]{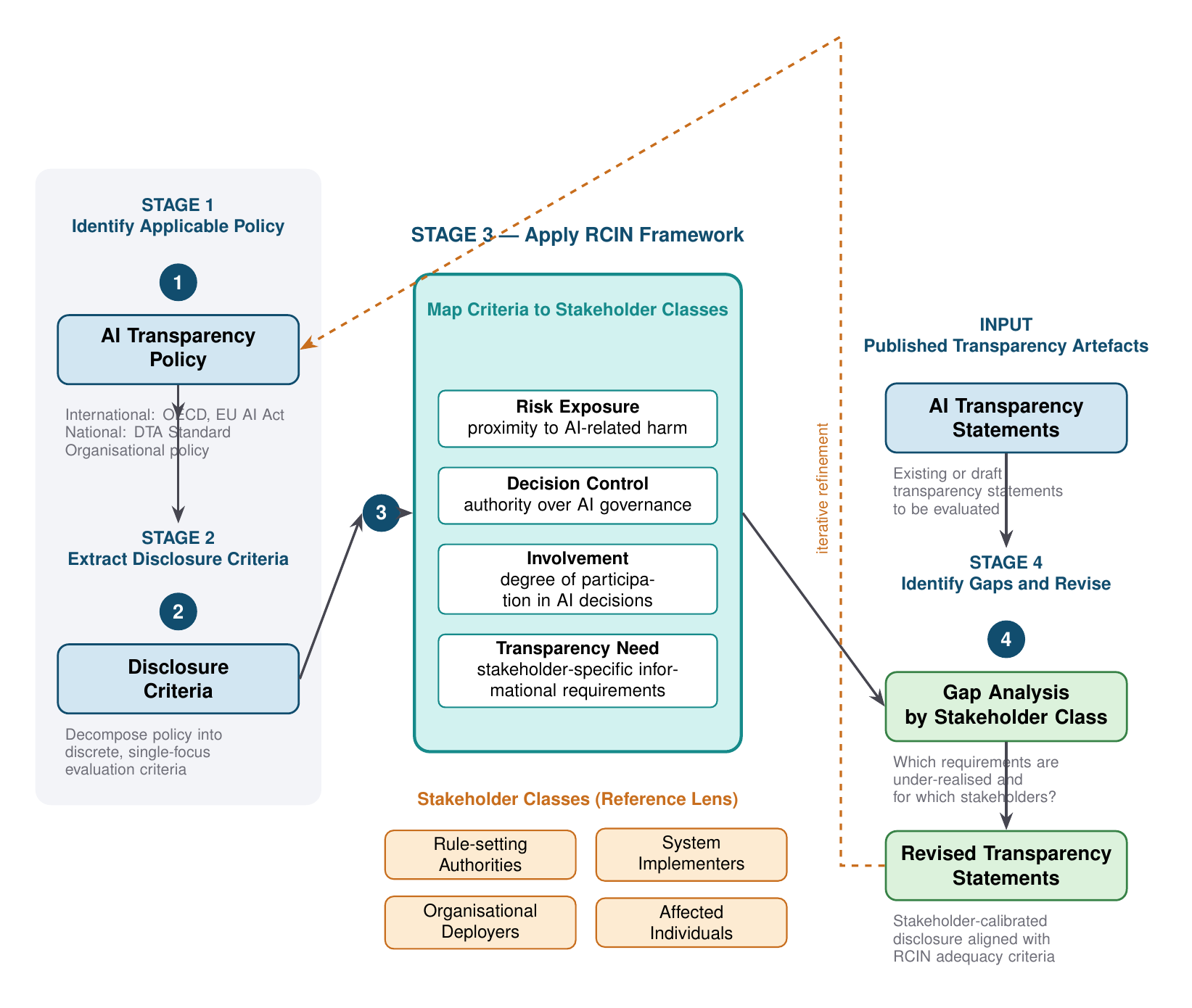}
\caption{RCIN framework to evaluate AI transparency artefacts.}
\label{fig:rcin}
\end{figure}

Within the RCIN framework, \emph{transparency adequacy} is defined as the degree to 
which transparency mechanisms are calibrated to stakeholder-specific needs given 
their risk exposure, decision control, and involvement. Misalignment occurs when 
artefacts are calibrated to high-control stakeholders while under-serving those who 
bear the greatest risk but exercise the least influence over transparency design. The 
structural profiles of each class are summarised in Table~\ref{tab:stakeholder_profile}, 
which serves as the analytical lens for the empirical analysis that follows. Figure~\ref{fig:rcin} illustrates how the framework can be applied in practice to evaluate any governance-mandated AI transparency artefact against stakeholder-specific needs.

\begin{table*}[t]
\caption{Stakeholder Risk--Control--Involvement--Need (RCIN) Profile}
\label{tab:stakeholder_profile}
\centering
\footnotesize
\begin{tabular}{p{3.5cm} p{2.0cm} p{1.2cm} p{3.5cm} p{5.2cm}}
\toprule
\textbf{Stakeholder class} & \textbf{Risk} & \textbf{Control} & \textbf{Extent of Involvement} & \textbf{Transparency Need} \\
\midrule

Rule-setting authorities 
& Low--Moderate 
& High 
& High (oversight and specification authority) 
& Traceability, documentation completeness, compliance verification, auditability \\

System implementers 
& Moderate 
& High 
& High (design and implementation influence) 
& Specification clarity, technical explainability, operational traceability \\

Organisational deployers 
& Moderate  
& Moderate 
& Medium (operational governance participation) 
& Governance visibility, accountability mapping, risk management clarity \\

Affected individuals / Citizens
& High 
& Low 
& Low (primarily informational; limited design influence) 
& Accessible explanation of decision impact, clarity of rights, procedural recourse and contestability mechanisms \\

\bottomrule
\end{tabular}
\end{table*}

\section{Research Design}
\label{sec:method}

This study adopts a socio-technical evaluation design \cite{Baxter2011Sociotechnical} 
to examine how governance-mandated AI transparency requirements are structured and 
realised in practice. The objective in this paper is not to benchmark agencies or rank their transparency 
maturity, but to analyse how mandated disclosure criteria distribute informational 
obligations across stakeholder classes and whether this distribution aligns with the 
RCIN framework introduced in Section~\ref{sec:framework}. The design proceeds in 
three stages: (i) structural decomposition of the mandated transparency policy into 
discrete criteria; (ii) stakeholder mapping of those criteria through the RCIN 
framework; and (iii) examination of how criteria are realised across 
published transparency statements.

\subsection{AI Transparency Statements Dataset}

The dataset consists of publicly available AI transparency statements 
published by Australian Government agencies under Version 1.1 of the 
DTA policy for responsible use of AI in government (effective 1 September 2024). Data collection and rubric 
development were conducted in May 2025. At the time of collection, 100 agencies were within scope of the mandate. Of these, 92 had publicly accessible statements suitable for analysis, with the remaining 8 excluded due to statements not yet published, no current 
AI use declared, or classified operational context precluding public disclosure. Statements vary considerably in scope 
and operational detail, ranging from active AI deployments embedded in 
service delivery to exploratory or limited AI use. Each statement is 
treated as a governance artefact produced in response to externally 
specified disclosure requirements. A revised Version 2.0 of the policy took effect on 15 December 2025, introducing additional mandatory requirements including strategic AI 
adoption planning, use case accountability, and risk-based impact 
assessment \cite{DTA2025Transparency}. This study therefore analyses 
the first generation of published statements under the original mandate, 
providing a baseline against which the impact of policy revision can be 
assessed in future work. All statement URLs, the evaluation instrument, 
and consolidated scoring sheets are provided as a publicly accessible 
replication package via \cite{AITransparencySupplementary2026}.

\subsection{Decomposition of Mandated Transparency Criteria}

The DTA Standard specifies minimum disclosure requirements encompassing agency 
intention for AI use, classification of usage patterns and domains, disclosure of 
direct public interaction, monitoring and oversight arrangements, risk mitigation 
practices, legal and policy compliance, and accessibility requirements 
\cite{DTA2025Transparency}. These mandated elements were decomposed into 20 discrete evaluation criteria \cite{AITransparencySupplementary2026}, 
separating bundled governance expectations into single-focus items to reduce 
ambiguity and permit granular analysis. The criteria were organised into three 
thematic groups:

\begin{itemize}
    \item \textbf{AI Use and Planning} (Criteria 1A--3B): purpose and intent for 
    current and future AI use, usage pattern classification, domain identification, 
    direct public interaction, and human review arrangements.
    \item \textbf{Governance and Risk Management} (Criteria 4A--7B): performance 
    monitoring, formal oversight bodies, legal and regulatory compliance, negative 
    impact mitigation, and alignment with DTA policy and Australian AI ethical 
    principles.
    \item \textbf{Public Accountability and Accessibility} (Criteria 8--13B): 
    public publication and discoverability, publication and last-updated dates, 
    review cycle commitments, plain language clarity, public contact mechanisms, 
    and ease of navigation.
\end{itemize}

We also introduced two qualitative supplementary metrics to explore evaluator 
interpretation variability: an \emph{Ambiguity Score} (1--5, measuring language 
specificity and clarity) and a \emph{Sentiment Tone} classification (Open, Neutral, 
or Defensive, reflecting the statement's overall communicative posture). These were 
not used as primary analytical outcomes but examined whether evaluators with 
different interpretive perspectives would reach divergent assessments of the same 
disclosures, reflecting the RCIN framework's relational conception of transparency 
\cite{Chen2021INTRPRT, Langer2021What}.

\subsection{Stakeholder Mapping of Mandated Criteria}

Each criterion was mapped to the stakeholder class whose transparency need it 
primarily serves, based on functional intent rather than implementation quality:

\begin{itemize}
    \item Criteria relating to oversight bodies, legal compliance, and DTA policy 
    alignment (4B, 5, 7A) primarily serve \textbf{rule-setting authorities}, 
    supporting auditability and regulatory traceability.
    \item Criteria concerning usage pattern classification, domain identification, 
    and system purpose (1A, 2A, 2B) primarily serve \textbf{system implementers} 
    and \textbf{organisational deployers}, supporting operational clarity and 
    governance accountability.
    \item Criteria addressing direct public interaction, human review, risk 
    mitigation, plain language, and contact mechanisms (3A, 3B, 6, 11, 12) 
    primarily serve \textbf{affected individuals and citizens}, supporting 
    informational accessibility and contestability.
\end{itemize}

The full mapping is presented in Table~\ref{tab:rci_rubric_mapping} in 
Section~\ref{sec:results}. This stage enabled examination of the design of mandate
itself, that is, whether the mandate distributes transparency obligations proportionately 
across stakeholder classes given their risk exposure, decision control, and 
involvement.

\subsection{Empirical Realisation and Descriptive Scoring}

The 100 government agencies within scope were distributed across three evaluators for independent assessment. At the time of data collection, 92 agencies had publicly accessible AI transparency statements available for analysis, while the remaining agencies had not yet published a statement or reported no applicable AI deployment. Each evaluator assessed approximately 30--33 statements using the standardised evaluation instrument and scoring guide. Statements were evaluated on a three-point ordinal scale (0 = not addressed; 1 = partially addressed; 2 = clearly and fully addressed) based strictly on explicit textual disclosure. Scores were assigned without inference from organisational context, public reputation, or external knowledge of agency AI programmes: if a criterion was not addressed in the published statement, it was scored zero regardless of what the agency may have communicated elsewhere. Two pilot rounds refined the instrument prior to full deployment, with the first round clarifying criterion definitions and scope, and the second stabilising the scoring guide and answer scale. Following the independent evaluations, the second author (project lead) conducted structured reconciliation meetings with all three evaluators to review scores, discuss interpretive ambiguities, and compare qualitative observations across the full dataset. This process supported consistency in interpretation despite the absence of overlapping double-coding, and resulted in a consolidated evaluation sheet included in the replication package \cite{AITransparencySupplementary2026}.

The evaluation did not seek to rank agencies or benchmark individual performance: its purpose is to identify structural patterns in how transparency obligations are encoded and realised across the mandate as a whole. Qualitative observations and criterion-level notes were recorded for 76 of the 92 evaluated statements. The remaining statements contained limited disclosure detail beyond basic template completion, restricting meaningful qualitative interpretation. These notes document specific disclosure practices, borderline scoring decisions, and recurring patterns of partial or absent disclosure that the ordinal scale alone cannot fully capture. The qualitative record was not subject to formal thematic coding but was reviewed comparatively across evaluators to identify recurring implementation patterns and corroborate quantitative findings.

\subsection{Analytical Strategy}

The study operationalises the RCIN framework across three analytical stages, as presented in Figure~\ref{fig:rcin}. First, the structural distribution of mandated criteria across stakeholder classes is examined to assess whether the governance design proportionately encodes transparency obligations relative to stakeholder risk and control asymmetries. Second, criterion-level scoring results are analysed across the 92 transparency statements to identify which disclosure requirements are consistently realised and which are weakly addressed or omitted.
Third, the RCIN framework is used to evaluate \emph{transparency adequacy}: the degree to which published transparency statements are calibrated to the informational needs of each stakeholder class given their structural position.  Stakeholder calibration is explored by aggregating criterion scores according to stakeholder mapping and comparing realisation levels across classes. Criteria mapped to affected individuals are examined against those mapped to rule-setting authorities and implementers to assess whether transparency adequacy reproduces the power asymmetries identified in the RCIN framework. Qualitative observations and criterion-level notes are used to contextualise quantitative patterns, identifying recurring disclosure practices across the corpus.

\subsection{Reproducibility of the RCIN Framework}

To support reproducibility and practical application, the complete evaluation instrument, scoring guidance, stakeholder mappings, and dataset of analysed transparency statements are provided through the accompanying replication package \cite{AITransparencySupplementary2026}. The RCIN framework is intended not only as an analytical lens for transparency statements, but as a general stakeholder-oriented mechanism for evaluating Responsible AI requirements in governance-driven contexts. Application of the framework involves three broad steps. First, relevant stakeholder classes are identified according to their structural role within the AI governance ecosystem, including rule-setting authorities, system implementers, organisational deployers, and affected individuals. Second, each stakeholder class is analysed across the four RCIN dimensions: risk exposure, decision control, degree of involvement, and stakeholder-specific need. Third, transparency or other Responsible AI requirements are evaluated according to whether they proportionately support the needs of each stakeholder class relative to their structural position.

\section{Results}
\label{sec:results}

This section presents the findings based on our two research questions: how 
mandated transparency criteria are structured for relevant stakeholder classes (RQ1), 
and whether published AI transparency statements meet the informational needs of 
those classes as defined by the RCIN framework (RQ2). 

\subsection{RQ1: Stakeholder Alignment with Transparency Criteria}

The mapping of 20 rubric criteria to stakeholder classes, presented in 
Table~\ref{tab:rci_rubric_mapping}, reveals that the transparency mandate is not 
stakeholder-neutral. Criteria serving \textbf{rule-setting authorities} require 
formal, institutionally anchored disclosure: named oversight bodies, legislative 
references, alignment with the DTA's AI transparency policy, and documented monitoring 
mechanisms. These criteria require agencies to reference specific named instruments, bodies, and policies, making compliance objectively verifiable against a fixed institutional standard.  This verifiability is reflected in consistently high realisation rates: criterion 7A (alignment with DTA AI transparency policy) was fully addressed by 88\% of agencies 
and criterion 5 (legal and regulatory compliance) by 66\%. Criteria serving \textbf{system implementers and organisational deployers} focus on operational intelligibility, supporting internal governance accountability for AI system deployment without directly addressing the transparency needs of individuals exposed to AI-supported decisions.

Criteria serving \textbf{affected individuals and citizens}, the stakeholder class 
with the highest risk exposure and lowest decision control, are fewer in number. Accessibility-oriented criteria (11: plain language; 12: public contact) address the form and 
readability of the statement rather than its informational substance. 
Contestability-oriented criteria (3A: disclosure of direct public 
interaction with AI; 3B: human review prior to AI-influenced outcomes; 
6: explanation of how negative impacts are identified and mitigated), 
address the substantive requirements that affected individuals have to understand, question, or seek redress from AI-supported decisions. These criteria are the least consistently addressed in the mandate: they do not require agencies to specify decision pathways, articulate review triggers, or describe formal mechanisms through which affected individuals can contest outcomes. The mandate thus encodes a structural ceiling on the degree of procedural empowerment that affected 
individuals can derive from any published transparency statement, regardless of how 
well agencies implement it. This asymmetry reflects the governance dynamic identified 
in the RCIN framework: those who define transparency obligations are not those who 
bear the consequences of AI-supported decisions \cite{Wang2025Signaling, bano2018power}.

\begin{tcolorbox}[colback=gray!10,colframe=gray!40,title=\textbf{RQ1 Finding}]
The transparency mandate is not stakeholder-neutral. Criteria serving 
high-control stakeholders are institutionally specific, administratively 
verifiable, and consistently realised. Criteria serving affected 
individuals are fewer, require less substantive disclosure, and split 
between how the statement is presented and what individuals need to 
understand or contest AI-supported decisions. The latter are the least 
consistently realised, reflecting a structural gap that agency 
implementation alone cannot close.
\end{tcolorbox}

\subsection{RQ2: The Adequacy of Transparency Statements Across Stakeholder Classes}

Our analysis reveals that while structural compliance with mandated 
criteria is widespread, transparency adequacy is unevenly calibrated 
across stakeholder classes, with disclosure serving high-control 
stakeholders consistently more specific and substantive than 
disclosure serving those with the greatest risk exposure.

Criteria serving rule-setting authorities and implementers are realised with high 
consistency and institutional specificity. Agencies routinely name oversight bodies, 
reference the Privacy Act or DTA policy, and confirm alignment with the Australian 
AI Ethics Principles. This governance-oriented transparency is clear, verifiable, 
and directed at stakeholders with the authority and institutional knowledge to 
interpret and act upon it. The DTA's own transparency statement achieved a score of 38 out of 40, exemplifying what complete structural compliance looks like in practice.

For affected individuals, the picture is more complex. Nearly all statements are 
written in plain language (96\% scored 2 on criterion 11) and are publicly 
accessible (88\% scored 2 on criterion 8), achieving broad formal reach. However, 
the substantive requirements that affected individuals need to understand, question, 
or contest AI-supported decisions are far less consistently realised. Disclosure of 
direct public interaction with AI (criterion 3A) was absent in 27\% of applicable 
statements; human review prior to AI-influenced outcomes (criterion 3B) was 
unaddressed in 18\% of cases; and explanation of how negative impacts are identified 
and mitigated (criterion 6) was fully addressed by only 48\% of agencies. For a 
citizen seeking to understand whether an AI system influenced a decision affecting 
their access to services, benefits, or rights, criteria 3A, 3B, and 6 represent the 
minimum informational foundation for meaningful engagement. Their inconsistent 
realisation is not merely a compliance gap; it is a transparency inadequacy. Three recurring configurations were observed across the dataset:

\begin{enumerate}
    \item \textbf{Governance-Dominant Transparency:} Strong institutional governance 
    disclosure with limited procedural specificity for affected individuals. This is 
    the most prevalent configuration, reflecting the structural emphasis of the 
    mandate itself.
    \item \textbf{Principle-Oriented Transparency:} High-level ethical commitments 
    asserted as organisational values without describing the mechanisms through which 
    they are operationalised for specific AI deployments or affected individuals.
    \item \textbf{Stakeholder-Calibrated Transparency:} Explicit system description, 
    risk articulation, oversight clarity, and meaningful recourse pathways serving 
    multiple stakeholder classes. This configuration is comparatively rare; the 
    Australian Electoral Commission (40/40) and Department of Industry, Science and 
    Resources (39/40) exemplify it.
\end{enumerate}

The first two configurations dominate, with governance-dominant transparency 
especially prevalent among agencies with active and complex AI deployments, precisely 
where affected individuals' need for procedural transparency is greatest. The distributional pattern across the dataset crystallises the central 
argument of this study. Agencies that score strongly on governance-oriented 
criteria, (criterion 7A fully addressed by 88\% of agencies and criterion 5 
by 66\%), consistently under perform on contestability-oriented requirements 
serving affected individuals, with criterion 6 fully addressed by only 48\% 
and criterion 3A unaddressed by 27\%. As Table~\ref{tab:rci_rubric_mapping} 
illustrates, agencies demonstrate strong structural fidelity in governance 
disclosure without carrying that fidelity through to the requirements most 
critical for those bearing the greatest exposure to AI-supported decisions.

The dataset also surfaces two boundary conditions of artefact-driven 
transparency. Several agencies declared no current AI use, 
rendering criteria non-applicable, while a small number did not 
publish statements due to the sensitive or classified nature of their work, 
where public disclosure would conflict with confidentiality and security 
obligations. Both cases illustrate that documentation-centric mandates 
presuppose a degree of operational openness that not all agencies can 
satisfy. This is raising questions about whether alternative accountability 
mechanisms are needed for contexts that fall outside the reach of public 
artefact disclosure \cite{Wang2025Signaling}. 

Widespread compliance with accessibility requirements combined with inconsistent realisation of contestability-oriented requirements constitutes what we term the \emph{Transparency Illusion}: transparency appears satisfied through mandated artefacts, yet adequacy, defined as proportional calibration to stakeholders' risk, control, need, 
and involvement, remains structurally uneven. This illusion is not produced 
by poor implementation by the organisations but by a mandate that encodes a structural 
ceiling on what transparency statements can provide to high-risk, 
low-control stakeholders regardless of how diligently agencies comply 
\cite{Balasubramaniam2022Transparency, Wang2025Signaling}.

\begin{tcolorbox}[colback=gray!10,colframe=gray!40,title=\textbf{RQ2 Finding}]
Compliance is widespread but transparency adequacy is unevenly calibrated. 
Disclosure serving high-control stakeholders is consistently detailed and 
institutionally specific. Disclosure serving affected individuals is 
accessible in form but inconsistently realised where it matters most: 
direct public interaction, human oversight, and impact mitigation. 
Compliance with mandated artefacts does not ensure adequacy for those 
most exposed to AI-supported decisions. This is the Transparency Illusion.
\end{tcolorbox}

\begin{table*}[t]
\caption{Alignment of Mandated Transparency Criteria with Stakeholder RCIN Positions}
\label{tab:rci_rubric_mapping}
\centering
\scriptsize
\setlength{\tabcolsep}{3pt}
\renewcommand{\arraystretch}{0.85}
\begin{tabular}{p{2.2cm} p{1.6cm} p{3.8cm} p{1.4cm} p{1.4cm} p{5.0cm}}
\toprule
\textbf{Thematic Group} & \textbf{Criteria} & \textbf{Primary Stakeholder} & \textbf{Risk Exposure} & \textbf{Decision Control} & \textbf{Transparency Adequacy Implication} \\
\midrule

AI Use \& Planning
& 1A, 2A, 2B
& System implementers / org.\ deployers
& Moderate
& Moderate--High
& Operationally clear; supports internal governance and procurement accountability \\
& 1B
& All stakeholders
& Varies
& Varies
& Mixed realisation; future intent frequently absent or non-specific \\
& 3A, 3B
& Affected individuals / citizens
& High
& Low
& Inconsistently realised; limited procedural depth; insufficient for meaningful contestability \\
\midrule

Governance \& Risk Mgmt
& 4B, 5, 7A
& Rule-setting authorities
& Low--Moderate
& High
& High structural fidelity; institutionally verifiable; consistently realised \\
& 4A, 7B
& All stakeholders
& Varies
& Varies
& Mixed realisation; performance monitoring and ethics alignment underspecified \\
& 6
& Affected individuals / citizens
& High
& Low
& Inconsistently realised; impact mitigation rarely described with procedural specificity \\
\midrule

Public Accountability \& Accessibility
& 11, 12
& Affected individuals / citizens
& High
& Low
& Near-universally realised; addresses form of disclosure but not protective content \\
& 8, 10, 13A, 13B
& All stakeholders
& Varies
& Varies
& Generally well realised; discoverability (13A, 13B) inconsistent \\
& 9A, 9B
& All stakeholders
& Varies
& Varies
& Systematically neglected; temporal accountability absent in nearly one third of statements \\
\bottomrule
\end{tabular}
\end{table*}

\section{Discussion}
\label{sec:discussion}

\subsection{Transparency Compliance Without Calibration}

The level of structural compliance observed across the dataset is satisfactory: the majority of agencies within scope had published statements and most mandated criteria were addressed at least partially. However, compliance itself is insufficient as a proxy for transparency adequacy. Institutional theorists have long observed that organisations subject to external mandates may decouple formal compliance from actual practice, producing what Meyer and Rowan 
describe as ceremonial adoption \cite{Meyer1977Institutionalized}. Brunsson similarly identifies organisational hypocrisy as the gap between decision, talk, and action under conflicting institutional demands \cite{Brunsson1989Organization}. The Transparency Illusion is a specific manifestation of this dynamic: transparency statements function as 
institutional signals of responsible practice \cite{Wang2025Signaling} 
without necessarily providing the stakeholder-calibrated substance that 
responsible practice requires.

This illusion is not produced by poor organisational intent but by the 
validation logic embedded in the mandate itself. Prevailing governance 
frameworks treat transparency as a disclosure problem: how much information 
should be provided and in what form \cite{EUAIAct2024, NIST2023AI}. The 
more fundamental problem is calibration. Because transparency obligations 
are defined and evaluated by high-control stakeholders, compliance criteria 
naturally reflect their informational priorities: auditability, policy 
alignment, and governance traceability. As defined in the RCIN framework, 
transparency adequacy is the degree to which transparency mechanisms are 
calibrated to stakeholder-specific needs given their risk exposure, decision 
control, and involvement. The empirical findings demonstrate that this 
calibration is structurally uneven: when risk exposure is inversely 
distributed relative to decision control, as is the case in AI governance, 
uniform disclosure templates entrench this asymmetry rather than correct 
it \cite{Wang2025Signaling, Zowghi2025On}.

The central question therefore shifts from how much transparency is 
provided to whether it is calibrated to those who depend on it most. 
What calibrated disclosure would require that current artefacts 
consistently omit is instructive: whether AI output is advisory or 
decision-making, what triggers human review, what pathway exists to 
contest an outcome, and when the statement was last reviewed. None of 
these are technically demanding. Their consistent absence suggests 
inadequate template design is the main constraint rather than implementation capacity.

\subsection{Alternative Explanations and Their Limits}

One might argue that contestability-oriented criteria score lower not 
because of structural asymmetry in the mandate's design, but because 
they are operationally harder to address, requiring knowledge of system 
behaviour and risk profiles that may be uncertain or evolving at the 
time of publication. This explanation has some merit but is insufficient 
as a complete account. The pattern of lower realisation on 
contestability-oriented criteria is consistent across agencies of 
vastly different size, resource capacity, and AI maturity. If 
operational complexity were the primary driver, agencies with more 
mature and resource-intensive AI programs would systematically 
outperform smaller agencies, which the data does not support. Moreover, the contestability-oriented criteria in question: whether 
a system directly interacts with the public (3A), human review 
occurs before AI-influenced outcomes affect individuals (3B), and 
if negative impacts are identified and mitigated (6), do not 
require technical expertise to address in a transparency statement. 
They require an organisational decision about what to disclose and a 
commitment to disclosing it. Their weaker realisation therefore more 
plausibly reflects these requirements being treated as lower priority 
within compliance practice, rather than agencies being genuinely 
unable to address them.

\subsection{Transparency Adequacy and Requirements Evolution}

Nearly one third of agencies omitted publication dates (criterion 9A, 
32\%) and last-updated dates (criterion 9B, 31\%), despite these being 
among the most basic accountability mechanisms available to any reader. 
Without this information, neither citizens nor oversight bodies can 
determine whether disclosures remain current or whether described 
systems have materially changed. This points to a dimension the RCIN 
framework does not yet fully address: requirements evolution. In RE, 
requirements change management recognises that requirements specified 
at a point in time must be actively maintained as systems, contexts, 
and stakeholder needs evolve \cite{Villamizar2021Requirements, 
Clement2023XAIR}. Without such mechanisms, transparency artefacts 
become stale requirements documents: formally present but no longer 
valid representations of the systems they describe. The policy revision from Version 1.1 to Version 2.0 \cite{DTA2025Transparency} 
illustrates this dynamic directly. The updated standard now explicitly 
requires statements to be reviewed at least annually, when agencies 
make significant changes to their AI approach, and when any new factor 
materially affects accuracy. That these update obligations were absent 
from the original mandate, and that nearly a third of agencies omitted 
even basic date information under it, suggests transparency requirements 
were not yet treated as subject to change management in governance 
practice. Future extensions of the RCIN framework should therefore 
assess whether artefacts commit to periodic re-elicitation and 
re-validation of stakeholder needs across the AI system lifecycle, 
rather than treating publication as a terminal compliance event.

\subsection{Generalisability of the Structural Mechanism}

Although this study focuses on transparency, the framework is transferable to other governance-oriented AI requirements such as fairness, accountability, human oversight, explainability, and contestability. For example, the AI transparency statement of the Attorney-General’s Department \cite{AGDAITransparency2025} describes governance oversight and responsible AI commitments, but the RCIN framework additionally prompts evaluation of whether affected individuals are provided with actionable recourse pathways, procedural clarity, and meaningful explanation relative to their structural position. In high-risk public-sector contexts, effective transparency therefore requires not only technical explanation mechanisms, but also stakeholder-calibrated information enabling individuals to understand, question, or challenge AI-supported outcomes. The framework supports evaluation of whether governance artefacts substantively serve those most exposed to AI-mediated decision consequences.

Similarly, the empirical setting in this study is Australian public-sector AI governance, 
the structural mechanism is jurisdiction-agnostic. Across major 
governance regimes, transparency is demonstrated through documentation 
artefacts where artefact completeness functions as the primary evidence 
of adequacy \cite{EUAIAct2024, NIST2023AI, OECD2019AI}, systematically 
privileging stakeholders positioned to interpret institutional artefacts 
without guaranteeing proportionally actionable transparency for high-risk, 
low-control stakeholders \cite{Balasubramaniam2022Transparency, 
McCormack2025comprehensive}. The structural asymmetry identified here is 
a predictable consequence of governance design in which transparency 
obligations are defined and audited by those insulated from their 
consequences \cite{Wang2025Signaling, Zowghi2025On}, and is therefore 
likely to recur across jurisdictions and sectors. The RCIN framework and the methodology presented in Figure~\ref{fig:rcin} 
are not bounded by the Australian context. Any organisation or regulatory 
body operating under an AI transparency mandate can apply the four-stage 
process to evaluate whether their transparency artefacts are 
proportionately calibrated to the needs of all relevant stakeholder 
classes.

\subsection{Recommendations for Requirements Engineering Practice}

When transparency requirements are specified by high-control stakeholders and 
validated through artefact compliance alone, the resulting requirements 
may systematically under-represent the needs of those most exposed to the 
consequences of AI-supported decisions. We offer four recommendations for RE research and practice.

First, transparency requirements should be decomposed into stakeholder-specific 
sub-requirements. Rather than specifying transparency as a single undifferentiated 
obligation, requirements elicitation should explicitly identify what each stakeholder class needs to understand, question, or act upon given their risk exposure, decision control, 
and degree of involvement \cite{Bell2023Think, Villamizar2021Requirements}. There are many requirements elicitation mechanisms that are suitable for this activity \cite{zowghi2005requirements}. Participatory approaches that directly involve representatives of affected individuals in transparency requirement specification are a productive direction for future work. The findings further suggest that explainability alone is insufficient to satisfy stakeholder-calibrated transparency requirements. While explanation mechanisms may help stakeholders understand how AI systems generate outputs or recommendations, many of the weaker criteria identified in this study concern governance visibility, accountability arrangements, contestability pathways, and human oversight rather than technical model interpretability alone. In governance-driven RE contexts, transparency therefore extends beyond explainability into a broader socio-technical requirement concerned with whether stakeholders receive information that is meaningful and actionable relative to their position, risk exposure, and ability to challenge AI-supported outcomes.

Second, current AI governance practice conflates \emph{satisfaction}, that is, conformance to specified criteria, with \emph{validation}, that is, confirmation that requirements fulfil stakeholder needs in context \cite{Mancine2024Estado, Ahmad2021Whats}. The evaluation instrument developed in this study offers an opportunity for stakeholder-oriented validation, though its further development and testing in high-risk domains is needed.

Third, transparency requirements, like all classes of requirements, should be treated as subject to change over time. As AI systems evolve, the transparency information relevant to each stakeholder class may change: new risks emerge, oversight arrangements are revised, and the degree of public interaction with AI systems may expand or contract. RE practice should therefore incorporate mechanisms for transparency requirement review across the AI system lifecycle, rather than treating published transparency statements as static artefacts. The widespread omission of publication and update dates identified in this study suggests that this dimension of requirements evolution is not yet embedded in governance practice. Fourth, transparency artefacts should be designed with explicitly specified acceptance criteria for each stakeholder class rather than as universal compliance documents \cite{Nannini2024Operationalizing, moreira2025explainable}. This does not require separate documents for each audience but does require that what constitutes sufficient disclosure for an affected individual be specified and assessed independently of what constitutes sufficient disclosure for a regulator, even within the same document.

\subsection{Practical Implications For Stakeholder Classes}

Beyond its theoretical contribution, we posit that the RCIN framework and its utilisation in this study offers practical value to each stakeholder class. For \textit{policy makers and rule-setting authorities}, it provides a diagnostic lens for evaluating the quality of the mandate, in particular, whether transparency mandates are structurally inclusive, making visible which stakeholder classes are proportionately served and which are under-served. Our findings further offer a concrete starting point: contestability-oriented requirements serving affected individuals are the least consistently realised in the current mandate and therefore the most productive target for policy revision. For \textit{system implementers (and requirements engineers)}, the framework translates abstract transparency principles into stakeholder-differentiated 
sub-requirements, providing a structured basis for decomposing transparency requirements
into specified, traceable obligations that integrate explainability, human 
oversight, and contestability mechanisms in a stakeholder specific information design.

For \textit{organisational deployers}, the framework functions as a validation guide for assessing whether published statements satisfy not only mandated criteria but the underlying stakeholder needs enabling proactive requirement refinement before statements are published rather than retrospective revision following audit. For \textit{affected individuals and citizens}, the framework advances a normative position: transparency requirements should be elicited through participatory processes that focus on risk exposure and informational need rather than derived from governance logic alone, supporting a shift from transparency as a compliance obligation toward transparency as a set of validated, stakeholder-calibrated requirements.

\section{Threats to Validity}
\label{sec:validity}

\textbf{Construct Validity.} Transparency is operationalised as formally encoded within governance-mandated public disclosure artefacts, deliberately delimited to evaluate transparency as a 
requirements realisation artefact within a specific governance regime. The evaluation instrument was derived from Version 1.1 of the DTA mandate, with each criterion mapped to stakeholder classes within the RCIN framework and scored strictly on explicit textual disclosure without inferring organisational intent or drawing on external knowledge of agency AI programmes. Alternative transparency models may yield different insights; the construct examined here is coherent with the governance regime under analysis 
and explicitly bounded as such. The stakeholder-to-criterion mappings within the RCIN framework were developed through theoretically grounded interpretive analysis by the authors rather than through direct stakeholder validation. While this enabled consistent structural analysis across the dataset, future work should incorporate participatory validation with affected stakeholders, regulators, and practitioners to further evaluate the robustness and transferability of these mappings.

\textbf{Internal Validity.} The 100 agencies within scope were divided across three evaluators, each responsible for approximately 33 agencies. However, at the time of analysis, 92 agencies had publicly accessible transparency statements suitable for evaluation. Because evaluators assessed distinct agency sets, traditional inter-rater reliability measures were not applicable. Interpretive ambiguity was minimised through discrete, single-focus criteria anchored to explicit disclosure indicators, a standardised scoring guide with worked examples, and two pilot rounds that refined criterion definitions and stabilised the scoring scale prior to full deployment. To support scoring and analysis consistency, the second author conducted structured reconciliation meetings with the evaluators to review scores, discuss interpretive ambiguities, and compare qualitative observations across the dataset. Although overlapping double-coding was not feasible within the scope of this study, this reconciliation process provided an additional layer of methodological consistency. Future replications could strengthen robustness through partial overlap sampling and formal inter-rater agreement analysis. Qualitative notes \cite{AITransparencySupplementary2026} recorded for 76 of 92 agencies further support scoring consistency: across evaluators, notes converge on identical criterion-level formulations, such as general ``mention of risk with no mitigation'' or ``no last updated date provided'', indicating that scoring was anchored to observable textual signals rather than interpretive judgement. The remaining 16 statements contained limited disclosure detail beyond basic template completion, restricting qualitative interpretation. The study does not attempt causal explanation for variation across individual agencies; it identifies recurring structural configurations across the corpus as a whole.

\textbf{External Validity.} The empirical setting is Australian public-sector AI transparency statements produced under a specific national mandate. Generalisation to other jurisdictions, sectors, or 
higher-risk deployment contexts should therefore be made cautiously. However, the structural mechanism identified, the divergence between artefact-level compliance and stakeholder-calibrated adequacy, is 
grounded in properties of documentation-centric governance that are common across contemporary AI regulatory regimes \cite{EUAIAct2024, NIST2023AI, OECD2019AI}. The subsequent revision 
of the DTA mandate to Version 2.0 (effective December 2025), which introduced contestability-oriented requirements absent from the original mandate, provides independent corroboration that the 
structural gaps identified in this study were recognisable at the policy level. The Transparency Illusion is proposed as an analytical construct describing this structural configuration rather than as a 
universal empirical claim.

\section{Conclusion and Future Directions}
\label{sec:conclusion}

Analysing 92 public-sector AI transparency statements 
based on the proposed RCIN framework, we find that formal compliance is widespread 
but proportional stakeholder calibration is uneven. Transparency is 
present, structured, and auditable, but not uniformly enabling for those 
most exposed to AI-supported decisions. This condition, the 
\emph{Transparency Illusion}, arises not from poor organisational practice 
but from a governance design in which those who define transparency mandates 
are not those who bear the consequences of AI-supported decisions. The RCIN 
framework provides a theoretically grounded lens for identifying this 
misalignment, and the evaluation instrument developed here offers one 
prototype for stakeholder-calibrated validation, though further development 
in higher-risk settings remains necessary. For RE, transparency must be engineered as a stakeholder-calibrated requirement rather than a single undifferentiated obligation. Validation must move beyond template verification toward proportional alignment. Transparency artefacts must carry explicit acceptance criteria for each stakeholder class rather than implicit assumptions of universal serviceability.

Future work should test the robustness of the Transparency Illusion across 
jurisdictions and higher-risk domains such as criminal justice, healthcare, 
and social services. Participatory methods for directly eliciting 
transparency requirements from affected individuals represent a significant 
gap in the current literature. Longitudinal investigation would address the 
temporal dimension identified in this study, and the development of 
stakeholder-differentiated acceptance criteria building on the RCIN 
framework offers a tractable research area for the RE community. As AI systems increasingly shape public outcomes, the critical question is no longer whether AI transparency exists but whether it meaningfully serves those who bear its consequences. 

\section{Data Availability Statement}

The replication package supporting this study, including the evaluation instrument, scoring guidance, stakeholder mappings, qualitative notes, and analysed transparency statements, is publicly available via Zenodo \cite{AITransparencySupplementary2026}.

\clearpage

\bibliographystyle{ieeetr}
\bibliography{References}

\vfill\eject

\appendices

\end{document}